\documentclass{appolb}
\usepackage{epsfig}% epsfig package included for placing EPS figures in the text
\begin{document}
\title{Studies of mesic nuclei via decay reactions
\thanks{Presented at International Symposium on "Mesic Nuclei", Krak\'{o}w, Sept. 20, 2013}}
%% you can use '\\' to break lines}
\author{S{\l}awomir Wycech
\address{National Centre  for Nuclear Studies, Ho\.za 69, 00-681 Warsaw,
Poland} 
\\
{Wojciech Krzemien}
\address{M. Smoluchowski Institute of Physics, Jagiellonian University, 30-059 Cracow, Poland}
}
\maketitle
\begin{abstract}
 Collisions in a system of two particles at
energies close to a bound state in different channels are
discussed. Next, the bound state decays into a third coupled
channel. A phenomenological approach to $dd\rightarrow \pi^-
p~^3$He reaction  is presented.
\end{abstract}

%\newpage
%\narrowtext
\section{Introduction}
\label{intro} The possibility of $\eta$-nuclear quasi-bound states
was first discussed by Haider and Liu  \cite{hai,bha} a long time
ago. The existence of such states has been elusive, however. At this
moment the experimental evidence is rather indirect, getting the most clear
indication from  the measurements of  $ pd \rightarrow \eta
^{3}$He~\cite{berger, mayer} and of $ dp \rightarrow \eta^{3}$He~\cite{timo,jurek-he3}
, and from the realization \cite{wil93}  that the rapid
slope of the cross section close to threshold may be a signal of
a quasi-bound state. The same behavior of the total-cross section was also confirmed in the photon induced reaction $\gamma ^3$He $\rightarrow \eta ^3$He~\cite{mami2}.  
The slope indicates large scattering length, but
the final state $\eta ^{3}$He interaction does not allow to
determine the sign of this length which would demonstrate  that
either  a bound state or a virtual state is observed.  Additional
information is necessary. One possibility is to use  the $
(\pi,\eta)$ reaction on a three-nucleon target.  Such an  analysis
indicates  that the $\eta ^{3}$He system is not a  bound but a
virtual state \cite{gw03}.

Analogical enhancement  close to the kinematic threshold was observed  in the total cross-section of the $dd \rightarrow ^4$He$ \eta$ reaction~\cite{Willis97,frascaria, wronska}. 
Again, these results suggest a large scattering length, however, do not give a conclusive answer whether the bound state exists.

Having the scattering length $A$ one may extrapolate the scattering
matrix 
\begin{equation}
\label{i1}
 T = \frac{A}{1-i p A},
\end{equation}
where $p$ is the $\eta-\mbox{He}$ relative momentum, at some distance below
the threshold. In this region $p= \sqrt{2\mu E}$ becomes complex, $p=
i|p|$. For Real $A>0$ one obtains the zero of the denominator and the
singularity of the $T$ matrix on the physical sheet. That means a
bound state for which  Im $A \neq0$ becomes  a quasi-bound state. In
case of Real $A<0$ the zero of denominator may also happen but for
$p=- i|p|$, it lays on the second Riemann sheet of the complex energy
plane. Such a state is called virtual and makes an analogy of
nucleon-nucleon, spin 0,  isospin 1 state known as anti-deuteron (
named so because of opposite sign of the pole position in the Im $p$
axis ). Going some distance below the threshold ( usually a short
distance as $A$ depends on energy and equation~\ref{i1} with
constant $A$  looses its  applicability)one may notice different
behavior of $ |T| $ in both cases. For a bound or quasi-bound state
$ |T| $ grows  up until the energy of the bound state is reached. On
the other hand, in case of a virtual state  $ |T| $ drops down
immediately below the threshold.

Direct observation of the elastic scattering amplitude below the
threshold is not feasible. However, one can observe a similar behavior
in channels  coupled to the channel where the bound state is
suspected to exist. Thus in the case of reaction
\begin{equation}
  \label{i2} dd\rightarrow \pi^-~p~^3\mbox{He}
\end{equation}
the channel of interest is
$\eta~^4$He and the decay channel consists of three particles
$\pi,p$ and $^3\mbox{He}$.

As the  $\eta~^3$He system seems almost bound,  the $\eta~^4$He
system is likely to be bound. The $\pi^-~p~^3$He  might be expected
to be the dominant decay channel. In such circumstances one could
expect a subthreshold  enhancement in the cross section for reaction
(\ref{i2}). Surprisingly, there is no experimental confirmation  of
such an effect. Measurements~\cite{sku12,adl13, krz13bis, krz13}, offer a
cross section of about 200 $nb$ which is apparently due to a
quasi-free reaction. An  upper limit of the fraction  that proceeds
through a quasi-bound state is obtained at a level of 25 $nb$~\cite{adl13}.

The aim of this work is to calculate/estimate  the magnitude of
\begin{equation}
  \label{i3} dd\rightarrow (\eta~^4\mbox{He})_{bound}\rightarrow \pi^-~p~^3\mbox{He}
\end{equation}
reactions and to offer some speculations on the existence of the
$(\eta~^4He)_{bound}$ state.

\section{Cross sections for the bound state formation and decays}

\subsection{Approximate amplitude for a two body process}
 \label{amplitude}

Consider transition  of  two initial  particles denoted by $ D , D'
$  into two particles  $B,B'$    of  a higher mass threshold.
Particles $BB'$ are assumed to form an unstable, $S$-wave,  bound
state $|B>$ of energy $E_B$ and width  $ \Gamma$. There may be
several modes of decay of this bound state and corresponding partial
widths are denoted by $ \Gamma_i$.

The reaction of interest consists of three steps

$\bullet~$ Colliding $ D , D'$ particles generate unstable state
$|B>$,

$\bullet~\bullet~$ Unstable state $|B>$ lives for some time and

$\bullet~\bullet~\bullet~$ the unstable state $|B>$ decays into
state $|F>$.

In this section all these stages of reaction are described in a
phenomenological way. The formulation used below is general but some
approximations are made for a specific case : $ D=D' = $ deuteron, $
B = ^4$He and $ B' = \eta$.

$\bullet~$It is assumed that the basic initial reaction
\begin{equation}
D + D' \rightarrow B + B'
 \label{c1}
\end{equation}
has been studied experimentally in some region above the $BB'$
threshold. The relevant cross section $\sigma_{DB}$ in the threshold
region may be presented in the form
\begin{equation}
\sigma_{DB} =~S(p_B)~p_B
 \label{c2}
\end{equation}
where the threshold behavior is described in part by $p_B$ - the
relative momentum in the $BB'$ channel. The function $S(p_B)$ is to be
extracted from experiment. With deeply bound or broad states
$S(p_B)$ is a weakly energy dependent function, for weak binding it
 may indicate a sharp threshold peak.

This cross section is generated by an operator $ V_{DB}$  which in a
standard way  allows to calculate the related scattering amplitude $
f_{DB}$
\begin{equation}
f_{DB} = \frac{2\mu_{BB'}}{4\pi} < DD|V_{DB}|B, p_B >.
 \label{c3}
\end{equation}
$\mu_{BB'}$ is the reduced mass and $p_B$,$p_D$ are the
relative momenta in the corresponding channels. The cross section
becomes
\begin{equation}
\frac{d\sigma_{DB}}{d\Omega}=  |f_{DB}|^2~\frac {p_B }{\mu_{BB'}}~
\frac {\mu_{DD}}{p_D }. \label{c4}
\end{equation}
A difficulty arises at this stage : from the scattering experiments
one can extract  $ | < DD|V_{DB}|B, p_B >|$  which is the modulus of
the  on-shell transition amplitude  for a given momentum $p_B$ while
one needs the transition to the bound state $ < DD|V_{DB}|B, E_B >
$. Formally
\begin{equation}
< DD|V_{DB}|B, E_B > =\int  d \textbf{p}_B  < DD|V_{DB}|B, p_B ><
p_B,B|B, E_B> \label{c5}
\end{equation}
where $< p_B,B|B, E_B>= \Psi_{BB'}(p_B)$ is the wave function of the
bound state in the momentum space. Equation (\ref{c5}) involves
integration over all momenta $p_B$ and not only over the momenta
allowed by the energy conservation.  To proceed without a  specific
model of $ V_{DB}$ we assume that the spacial range of this operator
is characterized by the size of the final particle B ( that is
$^4$He in the case of interest). This is the basic approximation of
this calculation,
\begin{equation}
< DD|V_{DB}|B ,p_B > =   |C_{DB}| \Psi_{B}(p_B) \label{c6}
\end{equation}
where $ \Psi_{B}(p_B)$ is the profile of single nucleon wave
function in nucleus $B$ folded over $\eta-$nucleon interaction range
and $|C_{DB}|$  is a constant determined from the slope of the cross
section.  For an estimate we use a gaussian
\begin{equation}
 \Psi_{B}(p)  = \exp (-\frac{R_B^2p^2}{2})
 \label{c7}
\end{equation}
and to simplify the estimate we assume the bound state wave function
in the same form
\begin{equation}
 \Psi_{BB'}(p)  = \exp (-\frac{R_{BB'} ^2p^2}{2})
~\left[ \frac{R_{BB'}^2}{\pi} \right]^{3/4}
 \label{c7bis}
\end{equation}

$\bullet~\bullet~$  The propagation of the bound state is described
by
\begin{equation}
 G_{BB'}  =   \frac{\Psi_{BB'}^*(p)\Psi_{BB'}(p)}{E- E_B+i\Gamma/2}
  \label{c8}
\end{equation}
where the complex part of the energy corresponds to the total decay rate.

$\bullet ~\bullet ~\bullet ~$  Decay of the  $|B>$  state into final
$|F_i>$ state is given by an operator $ V_{BF} $.  The matrix element of
this operator between the bound and the final state $<B| V_{BF}|Fi>
$   determines the partial  width of the state. The Fermi formula
gives
\begin{equation}
 \frac{\Gamma_i}{2}  = 2\pi   \int d \textbf{p}  |<B|
V_{BF}|F_i>|^2 \delta (E- E_F(p)) = (4\pi)^2 p_F~\mu_{FF'}
|<B|V_{BF}|F_i,p_F>|^2
  \label{c9}
\end{equation}
from which we obtain 
\begin{equation}
 |<B|V_{BF}|F_i,p_F>|^2 = \frac{\Gamma_i}{ 2p_F\mu_{FF'}(4\pi)^2}
  \label{c10}.
\end{equation}
In this calculation it is assumed that $^3$He is a spectator in the
decay process and the final decay energy is carried by the meson and
the proton. For simplicity the  non-relativistic phase space is
used. This may look suspicious in the $\pi$ meson case but  the
relevant reduced mass drops out from the final expression of the
cross section. This calculation may be easily improved anyway. Here
it serves  also as a check of the normalization used.

\subsection{Estimates of the  cross section}

The transition matrix element for the process in question is given
by
\begin{equation}
 <D| V_{D\rightarrow B\rightarrow F} |F_i> =  \int d\textbf{q} <D|V_{DB}|B q> \frac{<q|B,E_B> < B,E_B|
 V_{BF}|F_i>}
 {E- E_B+i\Gamma/2}
 \label{d1}.
\end{equation}
The related scattering amplitude is as in equation (\ref{c3})
\begin{equation}
 f_{D\rightarrow B\rightarrow F}  =  \frac{2\mu_{FF'}}{4\pi}   <D| V_{D\rightarrow B\rightarrow F} |F_i>
 \label{d2}.
\end{equation}
and the cross section
\begin{equation}
\frac{d\sigma_{DB}}{d\Omega}=  |f_{DB}|^2~\frac {p_B }{\mu_{BB'}}~
\frac {\mu_{FF}}{p_F } \label{d3}
\end{equation}

Collecting all factors and approximations the formula
\begin{equation}
\sigma_{DF}=  \frac{\sigma_{DB}}{p_B}~\frac {\sqrt{\pi}}{16}~
\frac{\Gamma^{\pi} } {( E- E_B)^2 +(\Gamma/2)^2}
\frac{1}{\mu_{BB'} R^3}
\end{equation}
is obtained where the two radii  were set equal $ R_B = R_{BB'}
\equiv R$.

 The factor $ \frac{\sigma_{DB}}{p_B}\mapsto ~C_{BD} \simeq0.3 nb /(MeV/c) $
may be obtained from experimental cross sections measured and
collected  in ref.\cite{bud09a}. With the expected values
$\Gamma^{\pi}\simeq 10~ MeV $, $\Gamma\simeq 20~ MeV $ and $
R\simeq2.5~fm $ one obtains $\sigma_{DF}\simeq 4.5~ nb$ at the peak.
The result is most sensitive to the radius $R$ but the numbers obtained
are below the experimental limit.  However, the relation to the
actual experimental limit is not that straightforward. It is  the
interference of reaction~\ref{i3} with the quasi-free reaction that
generates the experimental  limit of 25 $nb$. This requires
specific models and phase relations.  At this moment the question :

\section{Are there $\eta~^4$He bound states?}

cannot be fully answered neither by experiment nor by theory.
Simple, old  calculations of the threshold behavior in $\eta~^3$He
and $\eta~^4$He systems \cite{gwn95} indicated a  $\eta~^4$He bound
state. Since then  two basic ingredients have changed :

 $\bullet$  better  understanding of two nucleon  $\eta~NN \rightarrow N
 N$ decay mode.

$\bullet$  a better knowledge of the subthreshold $\eta N$ scattering
amplitude.

\begin{figure}[htb]

\begin{center}
%\vspace{.0cm} \special{psfile=ca3.ps  voffset = -450  hoffset= 120
%angle=0 vscale=60 hscale=60}
\includegraphics[width=0.8\textwidth]{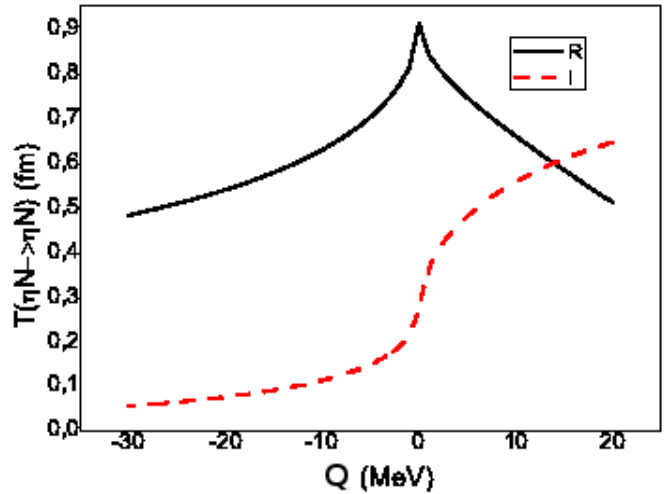}
\caption[F1] {The elastic $\eta$-N scattering amplitude  plotted
against the C.M. kinetic energy Q. Real part - continuous line,
absorptive part - dashed line.}
 \label{fig1}
%\vspace{5.5cm}
\end{center}
\end{figure}

The latter is represented by a best fit to multi-channel scattering
data obtained in  ref.\cite{gw05} and plotted in figure 1. An
average energy  involved in the $\eta N$ center of mass amounts to -
36 MeV ( 21 MeV binding and about 14 MeV of the residual nucleus
recoil). So far below the threshold the absorptive part  of the
amplitude is fairly small and the rate of $\pi^- p $ decay might be
strongly reduced.  That reduces the chance of observation via the
reaction (\ref{i3}). This plot shows also that the attractive
nuclear potential related to  Re $T_{\eta N}$ may be weaker than  in
the $\eta~^3$He case which involves  about $ -12$ MeV subthreshold
extrapolation.

On the other hand, the decay of the  $\eta~^4$He bound state into
two nucleon mode may be strongly enhanced.
 A  phenomenological evaluation of
the rates is possible as the cross sections for
\begin{equation}
\label{r2}
  p p \rightarrow p p \eta
\end{equation}
\begin{equation}
\label{r3d}
 p n  \rightarrow d \eta
\end{equation}
\begin{equation}
\label{r3}
 p n  \rightarrow  p n \eta
\end{equation}
have been  measured in the close to threshold region~\cite{czy07,mos10}. The analysis
based on the detailed balance corrected for final state interaction
has been performed in ref. \cite{kul98}. At central nuclear
densities the  related  absorptive potential of the $ \rho(r)^2 $
profile with a  strength Im $ W_{NN}( r=0) = 3.2 $ MeV was obtained.
However, Helium nucleus is twice as dense  and  the corresponding
absorptive potential rises to Im $ W_{NN}( r=0) \simeq13 $ MeV. Such
strong absorption may prevent binding or lead to much larger level
width. To resolve some of the problems it would be useful to have
also measurements of another
\begin{equation}
  \label{i4} DD\rightarrow (\eta~^4\mbox{He})_{bound}\rightarrow~p~n~D
\end{equation}
decay process.

From the experimental field, the ongoing analysis of the reactions $dd \rightarrow {^3\mbox{He}} p \pi^-$  and $dd \rightarrow {^3\mbox{He}} n \pi^{0} \rightarrow$ $^{3}\hspace{-0.03cm}\mbox{He} n \gamma \gamma$ from WASA-at-COSY, which will reach the sensitivity of several nb~\cite{krz13}, should help to answer the question of the existence of the bound state.

%uncomment the following lines to place a figure
%\begin{figure}[htb]
%\centerline{%
%\includegraphics[width=12.5cm]{Fig1}}
%\caption{Plot of ...}
%\label{Fig:F2H}
%\end{figure}

\vspace{0.5cm}

\noindent \emph{Acknowledgements} This work was supported by the the Polish National Science Center
under grants No. 2011/03/B/ST2/00270 and No. 2011/01/B/ST2/00431

\end{document}